\documentclass[
 aps,
 amsmath,amssymb,amsfig,
preprint,%
]{revtex4-1}

\usepackage{graphicx}
\usepackage{bm}
\usepackage{color}
\usepackage{dcolumn}
\usepackage{url}
\usepackage{multirow}

\usepackage[version=3]{mhchem} 

\usepackage{epstopdf}
\usepackage{subfigure}

\DeclareRobustCommand{\rchi}{{\mathpalette\irchi\relax}}
\newcommand{\irchi}[2]{\raisebox{\depth}{$#1\chi$}} 

\renewcommand{\vec}[1]{\boldsymbol{#1}}

\begin{document}

\title{Atomic orbital self-energy and electronegativity}

\author{M. Ribeiro, Jr.}
\email{ribeiro.jr@oorbit-us.com}
\affiliation{Office of Operational Research for Business Intelligence and Technology, USA}

\date{\today}

\begin{abstract}
  In this work, atomic calculations were performed within the local-density and generalized-gradient approximations of exchange and correlation density functionals within density-functional theory to provide accurate periodic trends of first ionization energies and electron affinities of the atomic series from hydrogen to xenon. Electronegativities were determined directly from Mulliken's formula and were shown to be equivalently calculated rather by using Slater-Janak's transition state or by calculating the electrostatic self-energies of the orbitals involved in the transition to ions. Finally, comparisons were made with other theoretical  and experimental results, including Mulliken-Jaff\'{e}'s electronegativity scale. 
\end{abstract}

\maketitle

\section{Introduction}

Electronegativity is the measure of an atom, molecule or solid substance to attract electrons to itself. The first connection of the electronegativity ($\rchi$) concept with quantum mechanics within density-functional theory (DFT) is assigned to Parr \textit{et al.}  \cite{Parr1978}, where $\rchi = -\mu = -(\frac{\partial E}{\partial N})_v$, where $\mu$ is the chemical potential in DFT \cite{Primer_in_DFT}. Slater claimed that the total energy ($E$) of an atom was a continuous function of its orbital occupations \cite{Slater_Book1974}. In a more general way one can describe the electronegativity of ions as

\begin{equation}
\rchi(Z,N) = -\frac{\partial E(Z,N)}{\partial N}, 
\label{eq:1}
\end{equation}

\noindent
for an ion with atomic number $Z$ and a variable number $N$ of electrons. If $Z=N$, then  Eq. \ref{eq:1} is valid for the neutral atom. By writing $E(Z,N)$ as a  parabolic function of $(N-Z)$ \cite{Hinze_JACS1963}, and choosing coefficients from atomic first ionization energy (I) and  electron affinity (A), we reach the following:

\begin{equation}
E(Z,N) = \frac{1}{2} (I-A)(N-Z)^2 - \frac{1}{2} (I+A)(N-Z) +E(Z,Z).
\label{eq:2}
\end{equation}

\noindent
Therefore, the partial derivative of Eq. \ref{eq:2} with respect to $N$, according to Eq. \ref{eq:1}, gives

\begin{equation}
\rchi(Z,Z) = \frac{(I+A)}{2}, 
\label{eq:3}
\end{equation}

\noindent
which is precisely Mulliken's electronegativity formula \cite{Mulliken_1934} with which I will work in this manuscript, with $I = E(Z, Z-1) - E(Z,Z)$ and $A = E(Z,Z) - E(Z,Z+1)$.

Usually, as a first approximation to the calculation of excitation energies, we simply take the difference between the ground-state Kohn-Sham eigenvalues. However, the Kohn-Sham eigenvalues (and wave functions) do not have any physical interpretation, excepting those related to the highest occupied state, which is regarded as the negative of the ionization energy of the system\footnote{Although the ionization energy calculated as the opposite of the eigenvalue of the highest occupied atomic orbital are far too small compared to experiment, almost by a factor of two. \cite{Primer_in_DFT_chapter6}} \cite{Almbladh_von_Barth_PRB1985}. Atomic ionization energies can be accurately calculated using DFT with many local or gradient types of exchange-correlation functionals, and so the atomic electronegativities \cite{Sen,Putz2005,Bartolotti_JACS1980,Robles_JACS1984}, given that total energy can be differentiable with respect to the number of electrons \cite{Perdew_PRL1982}. The electronegativity can be written in terms of the electron affinity and ionization energies of a system with $N = Z$ electrons as one-electron energy differences, supposing a maximum of two orbitals involved \cite{Sen}:

\begin{equation}
\label{eq:gap_definition}
\begin{split}
\rchi = \frac{1}{2}(I + A) &= \frac{1}{2}[E(Z, Z-1) - E(Z,Z)   +    E(Z,Z) - E(Z,Z+1)] \\
&= -\frac{1}{2}[E(Z, Z-1) - E(Z, Z+1)] \\
&= -\frac{1}{2}(\varepsilon_i + \varepsilon_j),
\end{split}
\end{equation}
   
\noindent
where $i$ and $j$ refer to atomic states which, for convenience, we label highest occupied atomic orbital (HOAO) and lowest unoccupied atomic  orbital (LUAO). This means one has to be able to calculate excited states to calculate electronegativities. However, DFT does not account for calculating excited states, as is widely known and described in many references elsewhere\cite{Primer_in_DFT,Perdew_PRL1983,Sham_PRL1983,Mori_PRL2008,Cohen_Science2008}. Therefore, the calculation of $LUAO-HOAO$ gap within DFT defined at a first approximation by $LUAO-HOAO = (LUAO-HOAO)^{KS} + \Delta_{xc}$ (not considering spin-orbit effects) is not accurately determined within DFT, as the exchange-correlation energy $\Delta_{xc}$ is only roughly approximated \cite{Primer_in_DFT}. 
In this work, theoretical results for periodic properties of ionization energy, electron affinity and electronegativity are presented using a method of fractional occupations of the orbitals involved in the transition from neutral atoms to ions ($\pm$0.5 electron, for negative/positive ions) using different exchange-correlation functionals within density-functional theory. The numerical fact of the (piecewise) linearity of the eigenvalue as a function of its occupation (at least for the range tried, of $\pm$0.5 electron, for which it proved valid from H to Xe) is used to calculate self-energy of the orbital(s) involved in the transition, for each element, and those same properties were determined and compared with the direct ion calculation. Details are described in next section.

\section{Methods of calculation}
\label{sec:method}
Electronegativity calculations by means of fractional occupations is not novelty, and was performed successfully several times \cite{Manoli_JCP1984,Robles_JACS1984,Bartolotti_JACS1980,Sen_JCP1981}. To rewrite the electronegativity expression in terms of half-occupations, let us consider a system - let's say, an atom - in its ground state to have a fully occupied valence orbital and unoccupied conduction orbital. If we write $\varepsilon(n_i,n_j)$ as the explicit dependence of the eigenvalue with the occupations at HOAO ($n_i$) and LUAO ($n_j$), then we can label $\varepsilon_j(1,1/2)$ the formerly LUAO of the atom in its ground state, $\varepsilon_i(1,0)$, that has just received half-electron. Similarly,  $\varepsilon_i(1/2,0)$ will be the formerly occupied valence state of the ground state system, $\varepsilon_i(1,0)$, that has been half-ionized. Assuming that the $LUAO-HOAO$ gap depends only on $i$ and $j$ orbitals occupations ($n_i$ and $n_j$), we can expand the energy in Taylor series around the ground state $(1,0)$ until second order:

\begin{equation}
\label{eq:E-expansion}
E(n_i,n_j)=E(1,0)+\frac{\partial E}{\partial n_i}n_i+
\frac{\partial E}{\partial n_j}n_j+
\frac{1}{2}\frac{\partial^2 E}{\partial n_i^2}n_i^2+
\frac{1}{2}\frac{\partial^2 E}{\partial n_j^2}n_j^2+
\frac{\partial^2 E}{\partial n_i\partial n_j}n_in_j,
\end{equation}

\noindent
with second derivatives with respect to total energies being regarded as the hardness tensor \cite{Chermette_CCR1998}. Now, Janak proved that the variation of total energy in DFT, with respect to the orbital occupation is equal to the eigenvalue of that orbital \cite{Janak}, 

\begin{equation}
\label{eq:Janak}
\partial E/\partial n_\alpha = \varepsilon_\alpha,
\end{equation}

\noindent
in which $E$ is the total energy and $n_\alpha$ is the occupation of the Kohn-Sham state $\alpha$ with eigenvalue $\varepsilon_\alpha$. Differentiating Eq. \ref{eq:Janak} a second time with respect to the orbital occupation, we obtain the relation $\partial\varepsilon_\alpha / \partial n_\alpha = 2S_\alpha$, in which $S_\alpha \propto \iint \frac{\rho_\alpha(\vec{r})\rho_\alpha(\vec{r'})}{|\vec{r} - \vec{r'}|}d\vec{r}d\vec{r'}$ is regarded as being the (electrostatic) self-energy function for the state $\alpha$ \cite{Guima, Mauro_ICPS2013, Guima_AIPADV2011} (\textit{v}. Appendix for details on this derivation). Writing

\begin{equation}\label{eq:2S}
\frac{\partial^2E}{\partial n_\alpha^2}=\frac{\partial \varepsilon_\alpha}{\partial n_\alpha}=2S_\alpha, 
\end{equation}

\noindent
with $\alpha=i,j$, we rewrite Eq. \ref{eq:E-expansion} as

\begin{equation}\label{eq:SvSc}
E(n_i,n_j)=E_0+\varepsilon_i(1,0)(n_i-1)+\varepsilon_j(1,0)n_j+
S_i (n_i-1)^2+
S_j n_j^2+
\frac{\partial^2 E}{\partial n_i\partial n_j}(n_i-1)n_j, 
\end{equation}

\noindent
where $E(1,0) = E_0$ is the ground state energy. Using Janak's theorem and differentiating Eq. \ref{eq:SvSc} in relation to $n_i$ and $n_j$ we obtain

\begin{subequations}
\begin{align}
\varepsilon_i(n_i,n_j)=\varepsilon_i(1,0)+ 2S_i (n_i-1)
+\frac{\partial^2 E}{\partial n_i\partial n_j}n_j\\
\varepsilon_j(n_i,n_j)=\varepsilon_j(1,0)+ 2S_j n_j
+\frac{\partial^2 E}{\partial n_i\partial n_j}(n_i-1).
\end{align}
\end{subequations}

So, we can calculate for half-ionized HOAO state, $\varepsilon_i(1/2,0)$,  and the half-occupied $j$ orbital (formerly LUAO orbital), $\varepsilon_j(1,1/2)$:

\begin{subequations}
\label{eq:eigenvalues_selfenergy}
\begin{align}
\varepsilon_i(1/2,0) &= \varepsilon_i(1,0)-S_i\\
\varepsilon_j(1,1/2) &= \varepsilon_j(1,0)+S_j. 
\end{align}
\end{subequations}

It is now possible to write $I$ and $A$ in terms of half-occupations for an atom with $N$ electrons as

\begin{subequations}
\label{eq:I_A_fractional}
\begin{align}
I &= \int_N^{N-1} \partial E = \int_{n_i=1}^{n_i=0} \varepsilon_i(n_i,n_j=0) \partial n_i = \varepsilon_i(0,0)-\varepsilon_i(1,0) = -\varepsilon_i(1/2,0)\\
A &= \int_{N+1}^N \partial E = \int_{n_j=1}^{n_j=0} \varepsilon_j(n_i=1,n_j) \partial n_j = \varepsilon_j(1,0)-\varepsilon_j(1,1) =  -\varepsilon_j(1,1/2).
\end{align}
\end{subequations}

Now, we can rewrite Eq. \ref{eq:gap_definition}, always considering the linearity of $\varepsilon_\alpha$ with occupation \cite{Slater_Wood_1971, *Goransson_PRB2005}, a necessary condition to be satisfied by exchange-correlation functionals\cite{Perdew_PRL1982}:

\begin{equation}
\label{eq:def_chi_fractional_occ}
\rchi = \frac{1}{2}(I + A) = -\frac{1}{2} [\varepsilon_j(1,1/2) + \varepsilon_i(1/2,0) ].
\end{equation}

Equation \ref{eq:def_chi_fractional_occ} can be rewritten in terms of integer occupations and self-energies using Eqs. \ref{eq:eigenvalues_selfenergy}:

\begin{equation}
\rchi = \frac{1}{2} (I + A) = -\frac{1}{2}(\varepsilon_i^{(0)} + \varepsilon_j^{(0)} + S_j - S_i).\label{eq:Eg-EgKS}
\end{equation}

\noindent
where I labelled $\varepsilon_i(1,0)$ and $\varepsilon_j(1,0)$ from Eq. \ref{eq:eigenvalues_selfenergy}  as $\varepsilon_j^{(0)}$ and $\varepsilon_i^{(0)}$, respectively. $S_i$ and $S_j$ can be calculated from the slope of the $\varepsilon_\alpha(n_\alpha)$  \textit{vs} $n_\alpha$ trend line. 

It is essential bearing in mind that there can be four situations for $\varepsilon_i$ and $\varepsilon_j$, which can be [a] $\varepsilon_i$ and $\varepsilon_j$ are orbitals of the same spin and same quantum numbers $n$ and $l$ (\textit{e.g.} Si, B, C); [b]  $\varepsilon_i$ and $\varepsilon_j$ are orbitals of different spins but same quantum numbers $n$ and $l$ (\textit{e.g.} N, P, As); [c] $\varepsilon_i$ and $\varepsilon_j$ are orbitals of the same spin, but different quantum numbers $n$ and $l$ (\textit{e.g.} Tc,Nb, Fe); and [d], $\varepsilon_i$ and $\varepsilon_j$ are orbitals of different spins and different quantum numbers $n$ and $l$ (e.g. Ar, Ti, Kr).

All the calculations were "all-electron" and performed with the ATOM program, contained in SIESTA package \cite{SIESTA}. The preferred exchange-correlation functional used was the GGA by Perdew, Burke and Ernzerhof (PB) \cite{PBE_PRL1996}. I encountered numerical convergence problems when trying to calculate electron affinities for some elements due to the prediction of unbound states\cite{Manoli_JCP1984},  specially when adding half-electrons to \textit{d}-states. Here, differently from  previously described by Robles and Bartolotti \cite{Robles_JACS1984}, where they calculate $A$ indirectly through $A = 2\rchi -I$, I simply changed the exchange-correlation functional to the BLYP functional \cite{Becke_PRA1988,*LYP_PRB1988}, or the LDA by Perdew and Wang (PW) \cite{LDA_PW}, and calculated $A$ directly. In the cases it also failed, as a common practice I then extrapolated the fitted trade line to $\varepsilon(n_\alpha)$ \textit{versus} $n_\alpha$ until $n_\alpha=0.5$ \cite{Primer_in_DFT_chapter6}. Although ionization energies were easily calculated for all elements using the PBE functional, for the final electronegativities table, I preferred to choose the best fit result regardless the exchange-correlation functional. I will comment this in more details in Results section.

\section{Results and discussion}

\begin{figure}[h!tbp]
\begin{center}
\includegraphics[width=16.5cm]{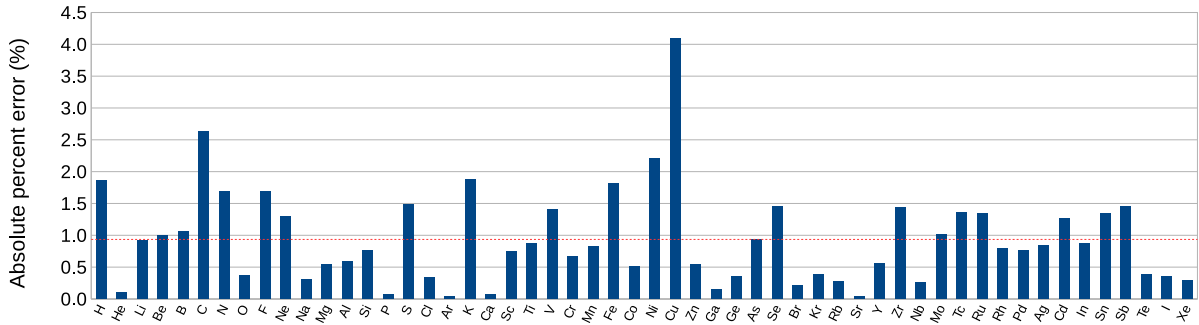}
\caption{\small{Plot of individual absolute percent errors to the experimental value of the first ionization energy calculated for each element from H to Xe. Red dotted line represents the average value of the absolute error in calculating I.}}
\label{fig:Figure1}
\end{center}
\end{figure}

Ionization energies are presented in Tab. \ref{tab:1}. Also indicated in Tab. \ref{tab:1} are the exchange-correlation functionals used for each element. It is worth mentioning that any single functional used (BLYP, PB and PW) provides very accurate results -- \textit{e.g.} an average percent error of less than 1.6\% if only PB is used --  but I decided to filter the best results out of the three functionals used for each element. Catches the eye the impressive agreement with experimental results, with an average error of just 0.9\% (\textit{v}. Fig. \ref{fig:Figure1}). 
Specially for heavier elements -- and for 3\textit{d} transition metals as well -- relativistic corrections (excluding spin-orbit) improved the results\cite{Rinaldo_JCP2008,Engel_PRB2001}.

\begin{table}
\centering
\tiny
  \caption{\tiny{Table showing the calculated first ionization energies from H to Xe. Columns headers represent: (1) Element symbol; (2) Exchange-correlation functional used in ionization energy calculation; (3) $\varepsilon_i$ represented by quantum numbers "n" and "l" of the orbital in which the calculation is done (\textit{v.} Eq. \ref{eq:I_A_fractional}a in text); (4) Orbital occupation; (5) First ionization energy, in electron-volts, calculated in present work; (6) Experimental first ionization energy, in electron-volts; (7) Absolute value of percent error in first ionization calculation if compared to experiment; (8) Absolute value of error in
 first ionization calculation if compared to experiment, in mili-electron-volts (meV). Values between angle brackets represent averages of the respective columns.}}
  \label{tab:1}
  \begin{tabular}{cccccccc}
    \hline
Element	&	XC 	&	nl	&	$n$	&	I 	&	I (exp)\cite{WEBELEMENTS}	&	$\mid\eta\mid$ (\%)	&	$\mid\eta\mid$ (meV)	\\
\hline
H	&	PW	&	1s	&	0.5	&	13.35	&	13.60	&1.85	&	252	\\
He	&	PW	&	1s	&	1.5	&	24.56	&	24.59	&0.11	&	26	\\
Li	&	PB	&	2s	&	0.5	&	5.34		&	5.39		&0.91	&	49	\\
Be	&	BLYP	&	2s	&	1.5	&	9.23		&	9.32		&0.99	&	93	\\
B	&	PB	&	2p	&	0.5	&	8.39		&	8.30		&1.06	&	88	\\
C	&	PB	&	2p	&	1.5	&	11.56	&	11.26	&2.63	&	296	\\
N	&	PB	&	2p	&	2.5	&	14.78	&	14.53	&1.69	&	246	\\
O	&	PB	&	2p	&	3.5	&	13.57	&	13.62	&0.36	&	49	\\
F	&	PB	&	2p	&	4.5	&	17.72	&	17.42	&1.69	&	294	\\
Ne	&	PB	&	2p	&	5.5	&	21.84	&	21.56	&1.29	&	278	\\
Na	&	PB	&	3s	&	0.5	&	5.12		&	5.14		&0.31	&	16	\\
Mg	&	PB	&	3s	&	1.5	&	7.60		&	7.65		&0.55	&	42	\\
Al	&	PW	&	3p	&	0.5	&	5.95		&	5.99		&0.58	&	35	\\
Si	&	PW	&	3p	&	1.5	&	8.21		&	8.15		&0.76	&	62	\\
P	&	PB	&	3p	&	2.5	&	10.49	&	10.49	&0.07	&	7	\\
S	&	PB	&	3p	&	3.5	&	10.21	&	10.36	&1.49	&	154	\\
Cl	&	PB	&	3p	&	4.5	&	13.01	&	12.97	&0.34	&	44	\\
Ar	&	PB	&	3p	&	5.5	&	15.75	&	15.76	&0.03	&	5	\\
K	&	PB	&	4s	&	0.5	&	4.42		&	4.34		&1.87	&	81	\\
Ca	&	BLYP	&	4s	&	1.5	&	6.11		&	6.11		&0.07	&	4	\\
Sc	&	PW	&	4s	&	1.5	&	6.61		&	6.56		&0.75	&	49	\\
Ti	&	PW	&	4s	&	1.5	&	6.89		&	6.83		&0.86	&	59	\\
V	&	PB	&	4s	&	1.5	&	6.84		&	6.75		&1.41	&	95	\\
Cr	&	BLYP	&	4s	&	0.5	&	6.81		&	6.77		&0.66	&	45	\\
Mn	&	PW	&	4s	&	1.5	&	7.50		&	7.43		&0.83	&	62	\\
Fe	&	PW	&	4s	&	1.5	&	8.05		&	7.90		&1.82	&	144	\\
Co	&	PB	&	4s	&	0.5	&	7.84		&	7.88		&0.51	&	40	\\
Ni	&	BLYP	&	4s	&	0.5	&	7.81		&	7.64		&2.21	&	169	\\
Cu	&	BLYP	&	4s	&	0.5	&	8.04		&	7.73		&4.10	&	316	\\
Zn	&	PB	&	4s	&	1.5	&	9.34		&	9.39		&0.55	&	51	\\
Ga	&	PW	&	4p	&	0.5	&	6.01		&	6.00		&0.15	&	9	\\
Ge	&	PB	&	4p	&	1.5	&	7.93		&	7.90		&0.35	&	28	\\
As	&	PB	&	4p	&	2.5	&	9.88		&	9.79		&0.94	&	92	\\
Se	&	PW	&	4p	&	3.5	&	9.89		&	9.75		&1.45	&	141	\\
Br	&	PB	&	4p	&	4.5	&	11.84	&	11.81	&0.22	&	26	\\
Kr	&	PB	&	4p	&	5.5	&	14.05	&	14.00	&0.38	&	53	\\
Rb	&	PB	&	5s	&	0.5	&	4.19		&	4.18		&0.28	&	12	\\
Sr	&	BLYP	&	5s	&	1.5	&	5.70		&	5.69		&0.03	&	2	\\
Y	&	PW	&	5s	&	1.5	&	6.25		&	6.22		&0.55	&	34	\\
Zr	&	PW	&	5s	&	1.5	&	6.54		&	6.63		&1.43	&	95	\\
Nb	&	PB	&	5s	&	0.5	&	6.78		&	6.76		&0.26	&	18	\\
Mo	&	PB	&	5s	&	0.5	&	7.02		&	7.09		&1.01	&	72	\\
Tc	&	PB	&	5s	&	0.5	&	7.18		&	7.28		&1.36	&	99	\\
Ru	&	PB	&	5s	&	0.5	&	7.26		&	7.36		&1.35	&	99	\\
Rh	&	PB	&	5s	&	0.5	&	7.40		&	7.46		&0.80	&	60	\\
Pd	&	PW	&	5s	&	1.5	&	8.27		&	8.34		&0.77	&	64	\\
Ag	&	PW	&	5s	&	0.5	&	7.64		&	7.58		&0.85	&	64	\\
Cd	&	BLYP	&	5s	&	1.5	&	9.11		&	8.99		&1.27	&	114	\\
In	&	PW	&	5p	&	0.5	&	5.74		&	5.79		&0.87	&	51	\\
Sn	&	PW	&	5p	&	1.5	&	7.44		&	7.34		&1.34	&	98	\\
Sb	&	PB	&	5p	&	2.5	&	8.73		&	8.61		&1.45	&	125	\\
Te	&	PW	&	5p	&	3.5	&	9.04		&	9.01		&0.39	&	35	\\
I	&	BLYP	&	5p	&	4.5	&	10.41	&	10.45	&0.36	&	37	\\
Xe	&	PB	&	5p	&	5.5	&	12.09	&	12.13	&0.29	&	35	\\
	&		&		&		&		 	&	        &$<$0.93$>$	&  $<$85$>$	\\
\hline
  \end{tabular}
\end{table}

\begin{table}
\centering
\tiny
  \caption{\tiny{Table showing the calculated electron affinity energies from H to Xe. Columns headers represent: (1) Element symbol; (2) Exchange-correlation functional used in electron affinity calculation; (3) $\varepsilon_j$ represented by quantum numbers "n" and "l" of the orbital in which the calculation is done (\textit{v.} Eq. \ref{eq:I_A_fractional}b in text); (4) Orbital occupation; (5) Electron affinity, in electron-volts, calculated in present work; (6) Experimental electron affinity, in electron-volts; (7) Absolute value of percent error in electron affinity calculation if compared to experiment; (8) Absolute value of error in electron affinity calculation if compared to experiment, in mili-electron-volts (meV); (9) Contribution of the calculated electron affinity to the electronegativity. Values between angle brackets represent averages of the respective columns. }}
  \label{tab:2}
  \begin{tabular}{ccccccccc}
    \hline
Element &	XC 	&	nl & $n$ & A 	&	A (exp)\cite{WEBELEMENTS}	&	$\mid\eta\mid$ (\%) 	&	$\mid\eta\mid$	(meV) &	Contrib. to EN (\%)	\\
H	&	PB	&	1s	&	1.50	&	0.75	&	0.75	&	0.09	&	1	&	5.14	\\
He	&	BLYP	&	2s	&	0.50	&	0.00	&	0.00	&	0.00	&	0	&	0.00	\\
Li	&	PW	&	2s	&	1.50	&	0.54	&	0.62	&	12.68	&	78	&	9.17	\\
Be	&	PB	&	2p	&	0.50	&	0.00	&	0.00	&	0.00	&	0	&	0.00	\\
B	&	PB	&	2p	&	1.50	&	0.16	&	0.28	&	41.97	&	117	&	1.90	\\
C	&	PB	&	2p	&	2.50	&	1.53	&	1.26	&	21.17	&	267	&	11.69	\\
N	&	PW	&	2p	&	3.50	&	0.06	&	0.07	&	20.21	&	15	&	0.39	\\
O	&	PB	&	2p	&	4.50	&	1.72	&	1.46	&	17.56	&	257	&	11.24	\\
F	&	PB	&	2p	&	5.50	&	3.73	&	3.40	&	9.57	&	325	&	17.38	\\
Ne	&	BLYP	&	3s	&	0.50	&	0.00	&	0.00	&	0.00	&	0	&	0.00	\\
Na	&	PW	&	3s	&	1.50	&	0.56	&	0.55	&	2.40	&	13	&	9.87	\\
Mg	&	PW	&	3p	&	0.50	&	0.30	&	0.22	&	35.84	&	78	&	3.75	\\
Al	&	PB	&	3p	&	1.50	&	0.39	&	0.43	&	10.58	&	46	&	6.13	\\
Si	&	PW	&	3p	&	2.50	&	1.47	&	1.39	&	5.47	&	76	&	15.13	\\
P	&	PB	&	3p	&	3.50	&	0.64	&	0.75	&	14.52	&	108	&	5.73	\\
S	&	PB	&	3p	&	4.50	&	2.15	&	2.08	&	3.30	&	68	&	17.37	\\
Cl	&	PB	&	3p	&	5.50	&	3.66	&	3.61	&	1.41	&	51	&	21.97	\\
Ar	&	BLYP	&	4s	&	0.50	&	0.00	&	0.00	&	0.00	&	0	&	0.00	\\
K	&	PB	&	4s	&	1.50	&	0.47	&	0.50	&	6.61	&	33	&	9.58	\\
Ca	&	BLYP	&	3d	&	0.50	&	0.01	&	0.02	&	76.68	&	19	&	0.09	\\
Sc	&	PB	&	3d	&	1.50	&	0.09	&	0.19	&	54.39	&	102	&	1.32	\\
Ti	&	BLYP	&	3d	&	2.50	&	0.09	&	0.08	&	5.11	&	4	&	1.31	\\
V	&	PB	&	3d	&	3.50	&	0.66	&	0.53	&	25.62	&	135	&	8.81	\\
Cr	&	PW	&	4s	&	1.50	&	0.62	&	0.68	&	8.52	&	58	&	8.28	\\
Mn	&	PW	&	3d	&	5.50	&	0.02	&	0.00	&	0.00	&	23	&	0.31	\\
Fe	&	PW	&	3d	&	7.50	&	0.16	&	0.16	&	0.53	&	1	&	2.08	\\
Co	&	PB	&	4s	&	1.50	&	0.98	&	0.66	&	47.89	&	318	&	11.12	\\
Ni	&	BLYP	&	4s	&	1.50	&	1.12	&	1.16	&	3.63	&	42	&	12.44	\\
Cu	&	PB	&	4s	&	1.50	&	1.16	&	1.24	&	6.17	&	76	&	12.55	\\
Zn	&	BLYP	&	4p	&	0.50	&	0.00	&	0.00	&	0.00	&	0	&	0.00	\\
Ga	&	PW	&	4p	&	1.50	&	0.46	&	0.43	&	8.09	&	35	&	7.35	\\
Ge	&	PB	&	4p	&	2.50	&	1.37	&	1.23	&	11.45	&	141	&	14.77	\\
As	&	PB	&	4p	&	3.50	&	0.68	&	0.80	&	15.13	&	122	&	6.47	\\
Se	&	PB	&	4p	&	4.50	&	2.10	&	2.02	&	3.97	&	80	&	18.09	\\
Br	&	PB	&	4p	&	5.50	&	3.47	&	3.36	&	3.23	&	109	&	22.68	\\
Kr	&	BLYP	&	5s	&	0.50	&	0.00	&	0.00	&	0.00	&	0	&	0.00	\\
Rb	&	PB	&	5s	&	1.50	&	0.46	&	0.49	&	4.92	&	24	&	9.93	\\
Sr	&	PW	&	4d	&	0.50	&	0.04	&	0.05	&	25.40	&	13	&	0.69	\\
Y	&	PW	&	4d	&	1.50	&	0.20	&	0.31	&	34.71	&	107	&	3.21	\\
Zr	&	PB	&	4d	&	2.50	&	0.40	&	0.43	&	5.25	&	22	&	5.89	\\
Nb	&	PB	&	4d	&	3.50	&	0.94	&	0.89	&	4.63	&	41	&	12.13	\\
Mo	&	PW	&	5s	&	1.50	&	0.66	&	0.75	&	11.54	&	86	&	8.61	\\
Tc	&	PB	&	4d	&	6.50	&	0.61	&	0.55	&	11.46	&	63	&	7.85	\\
Ru	&	PW	&	5s	&	1.50	&	1.04	&	1.05	&	0.23	&	2	&	12.57	\\
Rh	&	PW	&	5s	&	1.50	&	1.15	&	1.14	&	0.36	&	4	&	13.42	\\
Pd	&	PB	&	5s	&	0.50	&	0.50	&	0.56	&	10.72	&	60	&	5.98	\\
Ag	&	PW	&	5s	&	1.50	&	1.29	&	1.30	&	0.96	&	12	&	14.39	\\
Cd	&	PB	&	5p	&	0.50	&	0.00	&	0.00	&	0.00	&	0	&	0.00	\\
In	&	PB	&	5p	&	1.50	&	0.39	&	0.38	&	1.92	&	7	&	6.58	\\
Sn	&	PB	&	5p	&	2.50	&	1.23	&	1.11	&	10.25	&	114	&	14.59	\\
Sb	&	PW	&	5p	&	3.50	&	1.19	&	1.05	&	14.08	&	147	&	12.04	\\
Te	&	PB	&	5p	&	4.50	&	2.04	&	1.97	&	3.39	&	67	&	18.08	\\
I	&	PB	&	5p	&	5.50	&	3.10	&	3.06	&	1.34	&	41	&	22.37	\\
Xe	&	PB	&	6s	&	0.50	&	0.00	&	0.00	&	0.00	&	0	&	0.00	\\
	&		&		&		&		&		&	$<$11.39$>$&	$<$67$>$&$<$8.19$>$\\	

\hline
  \end{tabular}
\end{table}

\begin{figure}[h!tbp]
\begin{center}
\includegraphics[width=16.5cm]{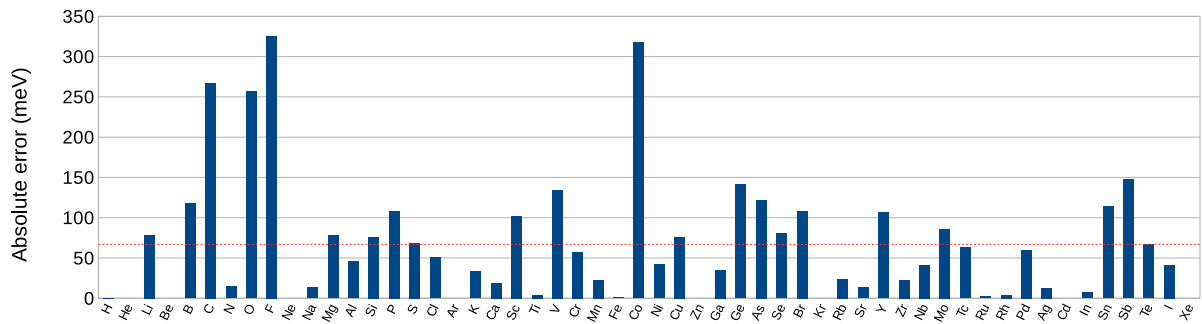}
\caption{\small{Individual absolute error, in eV, of the electro affinity calculated for each element from Z=1 (H) to Z=54 (Xe). Red dotted line represents the average value of the absolute error. }}
\label{fig:Figure2}
\end{center}
\end{figure}

\begin{figure}[h!tbp]
\begin{center}
\includegraphics[width=16.5cm]{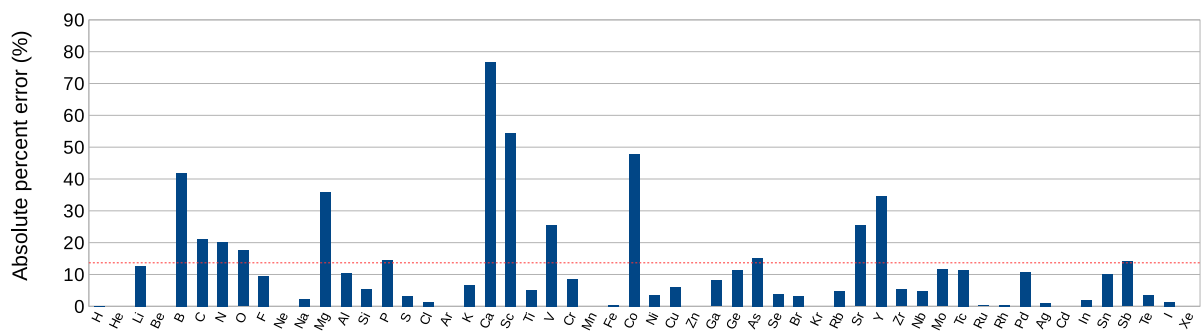}
\caption{\small{Individual absolute percent error of the electro affinity calculated for each element from Z=1 (H) to Z=54 (Xe). Red dotted line represents the average value of the absolute percent error.}}
\label{fig:Figure3}
\end{center}
\end{figure}

\begin{figure}[h!tbp]
\begin{center}
\includegraphics[width=16.5cm]{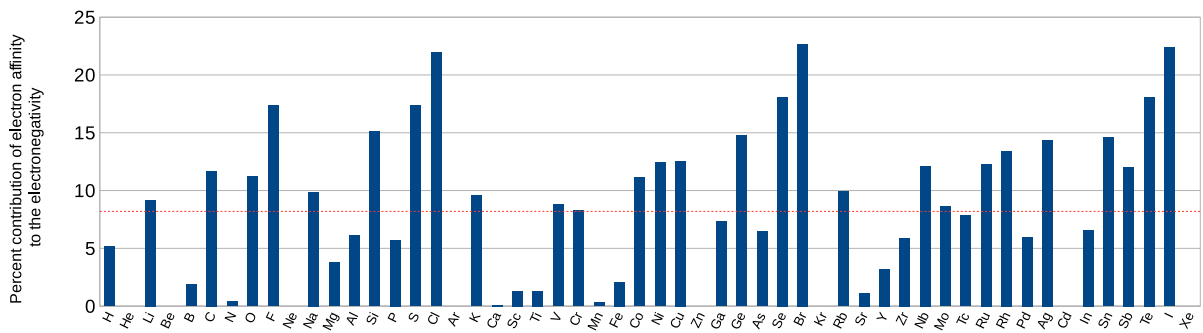}
\caption{\small{Contribution of the electro affinity of each individual element (from H to Xe) to its final electronegativity. Red dotted line represents the average value.}}
\label{fig:Figure4}
\end{center}
\end{figure}

The sign convention for electron  affinity was adopted as a positive energy value that indicates the spontaneous attachment of an electron to a single atomic orbital. Although first ionization energies are calculated with good accuracy using transition states\cite{Guima_AIPADV2011}, electron affinity still poses challenges from theoretical point of view, specially for numerical calculations within DFT\cite{Andersen2004,Sen,Lee_JPCL2010,Galbraith_JCP1996}, which traditionally found restrictions of applicability to this type of system due to large self-interaction errors that causes outermost occupied states to become unbound, specially in small ions\cite{Shore_PRB1977,Lee_JPCL2010,Galbraith_JCP1996}, thus making anions unstable upon electrons detachments as consequence of Koopman's theorem \cite{Kiracofe_2002}. 
In any case, I followed the tradition of, in principle, ignoring DFT drawbacks to calculate negative ions, and the electron affinity results are presented in Table \ref{tab:2}.
This time we can notice the much larger percent errors along the list in column 7, with the average percent error in calculations of 11.4\% in relation to experimental results. Some elements presented errors as large as 42\% (Boron), or even 77\% (Calcium). However, as the main objective in this work is to provide results for electronegativities, column 9 in Tab. \ref{tab:2} collects the contributions, to the electronegativity, of each individual electron affinity calculation. We can note that, for elements with the greatest electron affinity percent errors, like Ca, Mg and B, those electron affinity contribute little to the final electronegativities results (0.09\%, 3.75\% and 1,90\%, respectively). 

It is also important to mention the absolute errors in electron-volts. Typically, the LDA (spin-polarized) functional provides electron affinity results with absolute errors of the order of 0.3 to 1 eV (\textit{v}. Ref. \cite{Kiracofe_2002} and references therein). Comparing to present work, the average absolute error if merging the best results of each exchange-correlation functional is of only 0.07 eV, a considerable improvement. The average contribution of the electron affinity to the electronegativity is around 8.2\%, which indeed shows that the electron affinity value is not so important in final electronegativity value in Mulliken's formulation. Figures \ref{fig:Figure2}, \ref{fig:Figure3} and \ref{fig:Figure4} compile the results of Tab. \ref{tab:2}. Interestingly, in Fig. \ref{fig:Figure4} we can notice an increase in percent contribution of $A$ to $EN$ from left to right in periodic table, with peaks in F, Cl, Br and I. To summarize, sorting the best results among the exchange-correlation functionals used revealed the best strategy for DFT-based calculations, specially for electron affinity results, that can only be improved upon better exchange-correlation functionals, in which correlations play an important role \cite{Primer_in_DFT_chapter6}.

But the point here is not to provide the most accurate $I$, $A$ and $EN$ calculations, but to demonstrate that the transition state technique is still powerful in face of the modern exchange-correlation functionals, and that there is a connection of it with self-energies that could facilitate the numerical calculation and predict the correct trends for those atomic properties. Final electronegativities are grouped in Tab. \ref{tab:3}, together with some other DFT electronegativity scales like older results of Robles and Bartolotti\cite{Robles_JACS1984,Bartolotti_JACS1980} obtained by the $X_\alpha$ and spin-polarized $X_\alpha$ techniques, the Mulliken-Jaff\'{e} EN scale and the experimental results from direct applying $\rchi = \frac{1}{2}[I(exp)+A(exp)]$ (\textit{v}. Tabs. \ref{tab:1} and \ref{tab:2}).

\begin{table}
\centering
\tiny
  \caption{\scriptsize{Table showing the calculated electronegativities from H to Xe, all in electron-volts (eV). Columns headers represent: (1) Element symbol; (2) Electronegativity from X$\alpha$ calculations; (3) Electronegativity from spin-polarized X$\alpha$ calculations; (4) Electronegativities from Mulliken-Jaff\'{e}\cite{Putz2005}; (5) Experimental results from $\rchi = \frac{1}{2}(I+A)$;(6) Electronegativities from present work.}}
  \label{tab:3}
  \begin{tabular}{cccccc}
    \hline
Symbol	&	Xalpha\cite{Bartolotti_JACS1980}	&	Xalpha spin\cite{Robles_JACS1984}	&	Mulliken-Jaff\'{e}\cite{Putz2005}	&	Experimental\textsuperscript{\emph{a}}	&	Present work	\\
\hline
H	&	7.97	&	5.27	&	2.25	&	7.18	&	7.05	\\
He	&	12.61	&	7.93	&	3.49	&	12.29	&	12.28	\\
Li	&	2.58	&	1.69	&	3.1	&	3.00	&	2.94	\\
Be	&	3.8	&	3.52	&	4.8	&	4.66	&	4.62	\\
B	&	3.4	&	4.08	&	5.99	&	4.29	&	4.27	\\
C	&	5.13	&	6.39	&	7.98	&	6.26	&	6.54	\\
N	&	6.97	&	5.78	&	11.5	&	7.30	&	7.42	\\
O	&	8.92	&	6.45	&	15.25	&	7.54	&	7.64	\\
F	&	11	&	9.85	&	12.18	&	10.41	&	10.72	\\
Ne	&	10.31	&	6.6	&	13.29	&	10.78	&	10.92	\\
Na	&	2.32	&	1.67	&	2.8	&	2.84	&	2.84	\\
Mg	&	3.04	&	2.56	&	4.09	&	3.93	&	3.95	\\
Al	&	2.25	&	2.7	&	5.47	&	3.21	&	3.17	\\
Si	&	3.6	&	4.39	&	7.3	&	4.77	&	4.84	\\
P	&	5.01	&	4.38	&	8.9	&	5.62	&	5.57	\\
S	&	6.52	&	5.18	&	10.14	&	6.22	&	6.18	\\
Cl	&	8.11	&	7.5	&	9.38	&	8.29	&	8.34	\\
Ar	&	7.11	&	4.93	&	9.87	&	7.88	&	7.88	\\
K	&	1.92	&	1.47	&	2.9	&	2.42	&	2.45	\\
Ca	&	1.86	&	2.48	&	3.3	&	3.07	&	3.06	\\
Sc	&	2.52	&	3.4	&	4.66	&	3.37	&	3.35	\\
Ti	&	3.05	&	4.16	&	5.2	&	3.46	&	3.49	\\
V	&	3.33	&	4.09	&	5.47	&	3.64	&	3.75	\\
Cr	&	3.45	&	2.3	&	5.56	&	3.72	&	3.71	\\
Mn	&	4.33	&	3.38	&	5.23	&	3.72	&	3.76	\\
Fe	&	4.71	&	4.41	&	6.06	&	4.03	&	4.10	\\
Co	&	3.76	&	4.84	&	6.21	&	4.27	&	4.41	\\
Ni	&	3.86	&	5	&	6.3	&	4.40	&	4.46	\\
Cu	&	3.95	&	3.76	&	4.31	&	4.48	&	4.60	\\
Zn	&	3.66	&	3	&	4.71	&	4.70	&	4.67	\\
Ga	&	2.11	&	2.54	&	6.02	&	3.21	&	3.24	\\
Ge	&	3.37	&	4.1	&	8.07	&	4.57	&	4.65	\\
As	&	4.63	&	4.08	&	8.3	&	5.30	&	5.28	\\
Se	&	5.91	&	4.79	&	9.76	&	5.89	&	6.00	\\
Br	&	7.24	&	6.74	&	8.4	&	7.59	&	7.66	\\
Kr	&	6.18	&	4.36	&	8.86	&	7.00	&	7.03	\\
Rb	&	1.79	&	1.41	&	2.09	&	2.33	&	2.33	\\
Sr	&	1.75	&	1.98	&	3.14	&	2.87	&	2.87	\\
Y	&	2.25	&	2.59	&	4.25	&	3.26	&	3.23	\\
Zr	&	3.01	&	3.63	&	4.57	&	3.53	&	3.47	\\
Nb	&	3.26	&	2.3	&	5.38	&	3.83	&	3.86	\\
Mo	&	3.34	&	2.3	&	7.04	&	3.92	&	3.84	\\
Tc	&	4.58	&	3.72	&	6.27	&	3.91	&	3.90	\\
Ru	&	3.45	&	3.11	&	7.16	&	4.20	&	4.15	\\
Rh	&	3.49	&	3.23	&	7.4	&	4.30	&	4.27	\\
Pd	&	3.52	&	2.4	&	7.16	&	4.45	&	4.39	\\
Ag	&	3.55	&	3.39	&	6.36	&	4.44	&	4.47	\\
Cd	&	3.35	&	2.8	&	5.64	&	4.50	&	4.55	\\
In	&	2.09	&	2.48	&	5.28	&	3.09	&	3.06	\\
Sn	&	3.2	&	3.85	&	7.9	&	4.23	&	4.33	\\
Sb	&	4.27	&	3.84	&	8.48	&	4.83	&	4.96	\\
Te	&	5.35	&	4.43	&	9.66	&	5.49	&	5.54	\\
I	&	6.45	&	6.04	&	8.1	&	6.76	&	6.76	\\
Xe	&	5.36	&	3.85	&	7.76	&	6.06	&	6.05	\\
    \hline
  \end{tabular}\\
  \textsuperscript{\emph{a}} Calculated using Mulliken's formula with $I(exp)$ and $A(exp)$ from tables \ref{tab:1} and \ref{tab:2}. 
\end{table}

\begin{figure}[h!tbp]
\begin{center}
\includegraphics[width=16.5cm]{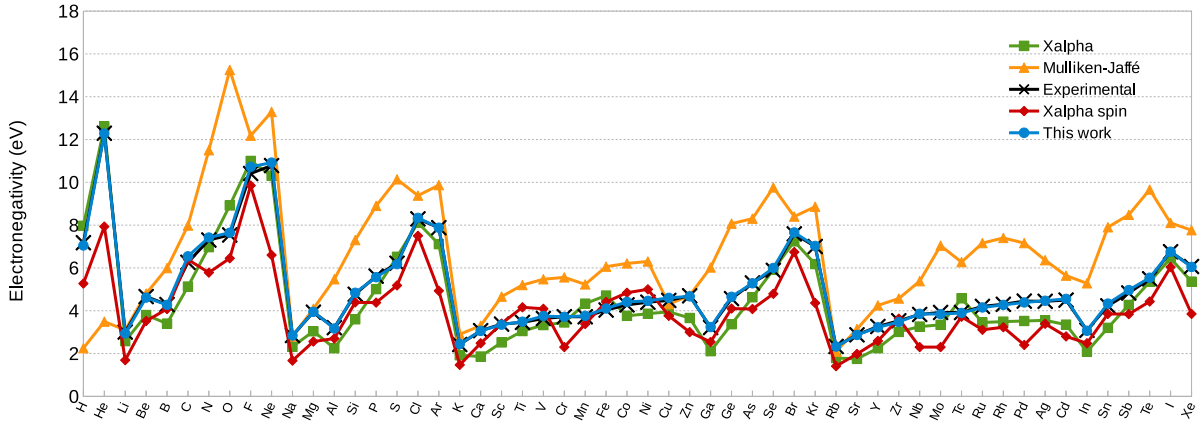}
\caption{\small{Electronegativity trends from H to Xe, comparing present results (blue/circles) to the old theoretical results from Bartolotti, Gadre and Parr\cite{Bartolotti_JACS1980} (green/squares) and Robles and Bartolotti\cite{Robles_JACS1984} (red/diamonds), as well as Mulliken-Jaff\'{e} data\cite{Putz2005} (orange/triangles) and experimental results (black/crosses) }}, 
\label{fig:Figure5}
\end{center}
\end{figure}

Figure \ref{fig:Figure5} summarizes the comparisons with $X_\alpha$ calculations and experiments. The agreement with experimental results represented a significant improvement, with an average percent error of only 1.2\%, compared to 16.3\% and 22.3\% of $X_\alpha$ and spin-polarized $X_\alpha$, respectively. Also, there is a clear improvement in describing periodic trends from K to Zn, including the 3\textit{d} transition metals, and from Rb to Cd, including 4\textit{d} transition metals.

The calculation of atomic orbital electrostatic self-energy is straightforward since the linearity condition of $\varepsilon(n_i)$ is valid to the extent of $\pm \frac{1}{2}$ electron, for the PW, PB and BLYP exchange-correlation functionals, for all elements from H (Z=1) to Xe (Z=54). Table \ref{tab:4} contains the results of self-energy calculations of a number of elements, obtained using Eq. \ref{eq:2S}. 
Electronegativity for some elements like Tc, Mn, Mg, Ne and Cd can be calculated with accurate results just by the first ionization energy, because of lack of numerical convergence and proven negative ion instability \cite{Andersen2004}. Apart from little numerical differences when considering the linear fit in different regions of the $\varepsilon(n_i)$ curve\cite{Perdew_PRL1982} (and as Eq. \ref{eq:eigenvalues_selfenergy} is an approximation to second order), self-energy-derived electronegativities results present excellent agreement with Tab. \ref{tab:3}.

\begin{table}
\centering
\scriptsize
  \caption{\small{Table showing the calculated electronegativities for some elements using self-energy results. Columns headers represent: (1) Element symbol; (2) Self-energy of the $i$ orbital (v. Table \ref{tab:1}); (3) Self-energy of the $j$ orbital (v. Table \ref{tab:2}); (4) Eigenvalue, for orbital $i$; (5) Eigenvalue for orbital $j$; (6) First ionization energy; (7) Absolute value of percent error in ionization energy; (8) Electron affinity; (9) Absolute value of percent error in electron affinity; (10) Electronegativity as calculated from self-energy; (11) Absolute value of percent error in electronegativity calculation from self-energy, if compared to electronegativity obtained by direct application of Mulliken's formula. Self-energies, ionization energies, electron affinities, eigenvalues and electronegativities in electron-volts (eV).}}
  \label{tab:4}
  \begin{tabular}{ccccccccccc}
    \hline
Element 	&	$2S_i$\textsuperscript{\emph{a}}	&	$2S_j$\textsuperscript{\emph{a}}	&	$\varepsilon_i$	&	$\varepsilon_j$ 	&	I &	$\mid\eta\mid$ in I (\%)	&	A	&	$\mid\eta\mid$ in A (\%)	&	EN \textsuperscript{\emph{b}}	&	$\mid\eta\mid$ in EN (\%)\textsuperscript{\emph{c}}	\\
\hline
Na	&	0.34	&	0.24	&	-0.22	&	-0.16	&	5.35	&	4.14	&	0.53	&	3.63	&	2.94	&	3.44	\\
Si	&	0.53	&	0.46	&	-0.34	&	-0.34	&	8.22	&	0.79	&	1.47	&	5.62	&	4.84	&	0.04	\\
P	&	0.62	&	0.45	&	-0.46	&	-0.27	&	10.49	&	0.04	&	0.67	&	10.36	&	5.58	&	0.26	\\
Cl	&	0.69	&	0.69	&	-0.60	&	-0.60	&	12.81	&	1.18	&	3.45	&	4.40	&	8.13	&	2.44	\\
Fe	&	0.45	&	0.79	&	-0.37	&	-0.40	&	8.05	&	1.85	&	0.13	&	22.66	&	4.09	&	0.42	\\
Co	&	0.46	&	0.37	&	-0.35	&	-0.25	&	7.84	&	0.47	&	0.97	&	45.81	&	4.41	&	0.13	\\
Ni	&	0.49	&	0.42	&	-0.31	&	-0.29	&	7.62	&	0.28	&	1.09	&	5.39	&	4.36	&	2.36	\\
Cu	&	0.50	&	0.40	&	-0.32	&	-0.29	&	7.77	&	0.62	&	1.17	&	5.17	&	4.47	&	2.79	\\
Ge	&	0.51	&	0.45	&	-0.33	&	-0.33	&	7.93	&	0.35	&	1.38	&	11.63	&	4.65	&	0.03	\\
Se	&	0.58	&	0.52	&	-0.44	&	-0.41	&	9.90	&	1.47	&	2.10	&	4.03	&	6.00	&	0.02	\\
Br	&	0.64	&	0.59	&	-0.55	&	-0.55	&	11.84	&	0.20	&	3.47	&	3.17	&	7.65	&	0.02	\\
Rb	&	0.26	&	0.21	&	-0.18	&	-0.14	&	4.19	&	0.33	&	0.45	&	6.48	&	2.32	&	0.12	\\
Ru	&	0.43	&	0.34	&	-0.32	&	-0.25	&	7.25	&	1.48	&	1.05	&	0.05	&	4.15	&	0.08	\\
Ag	&	0.46	&	0.38	&	-0.33	&	-0.29	&	7.64	&	0.89	&	1.29	&	1.31	&	4.47	&	0.01	\\
Te	&	0.51	&	0.46	&	-0.41	&	-0.39	&	9.05	&	0.41	&	2.12	&	7.75	&	5.58	&	0.79	\\
I	&	0.56	&	0.52	&	-0.49	&	-0.49	&	10.43	&	0.22	&	3.05	&	0.18	&	6.74	&	0.24	\\
    \hline
  \end{tabular}\\
\textsuperscript{\emph{a}} Electrostatic self-energy, in eV (\textit{v.} Eq. \ref{eq:2S} in text).\\
\textsuperscript{\emph{b}} Electronegativity, in eV, calculated by $S_i$ and $S_j$ (\textit{v.} Eq. \ref{eq:Eg-EgKS} in text).\\
\textsuperscript{\emph{c}}  Relative to the electronegativity calculated $0.5(I+A)$.
\end{table}

In summary, I calculated first ionization energies, electron affinities and electronegativities for elements ranging from H to Xe using  transition state techniques within density-functional theory using different exchange-correlation functionals. This method showed extremely accurate, with an average percent error of only 1.2\% compared to experimental results.  Numerical convergence was easily achieved for all $I$ calculations and most $A$ calculations as well, and alternatives were used to overcome convergence problems. A direct connection of electronegativity and atomic orbital self-energy was tested and used to endorse preliminary results, which showed excellent agreement with experimental ones, even without considering spin-orbit effects.

I believe this method can be employed with other computational techniques that include spin-orbit effects, like Weizmann theory \cite{Martin_1999}.

%
%


\newpage

\providecommand{\noopsort}[1]{}\providecommand{\singleletter}[1]{#1}%

\section*{Appendix}

Calculating the second derivative of $\partial E / \partial n_\alpha = \epsilon_\alpha$ (Eq. \ref{eq:Janak}),

\begin{equation}
\frac{\partial \epsilon_\alpha}{\partial n_\alpha} = \left\langle \psi_\alpha \left\vert \frac{\partial (T+V_H+v_{xc})}{\partial n_\alpha} \right\vert \psi_\alpha \right\rangle .
\label{apx:eq-derivada_Janak-01}
\end{equation}

Identifying terms:
\begin{itemize}

\item $ \left \langle \psi_\alpha \left \vert \frac{\partial T}{\partial n_\alpha} \right \vert \psi_\alpha \right \rangle = t_\alpha $

\item $V_H = \frac{\delta U}{\delta \rho}  = \int \frac{\rho(\vec r)}{|\vec r - \vec r'|} d\vec r $

\item $ \left \langle \psi_\alpha \left \vert \frac{\delta U}{\delta \rho} \right \vert \psi_\alpha \right \rangle = \int\rho_\alpha \frac{\delta U}{\delta \rho} d\vec r $

\item $\left \langle \psi_\alpha \left \vert \frac{\delta E_{xc}}{\delta \rho} \right \vert \psi_\alpha \right \rangle = \int\rho_\alpha \frac{\delta E_{xc}}{\delta \rho} d\vec r $

\item $ v_{xc} = \frac{\delta E_{xc}}{\delta \rho(\vec r)} $

\item $ V_H + v_{xc} = \int \frac{\sum_i n_i \rho_i(\vec r)}{|\vec r - \vec r'|}d\vec r' + \frac{\delta E_{xc}}{\delta \rho(\vec r)}  $

\end{itemize}

Calculating the partial derivative of $V_H + v_{xc}$ with relation to $n_\alpha$,

\begin{equation}
\begin{split}
\frac{\partial (V_H + v_{xc})}{\partial n_\alpha} &= \int \frac{\rho_i(\vec r)}{|\vec r - \vec r'|} d\vec r' + \int \frac{\sum_j n_j\frac{\partial \rho_j(\vec r)}{\partial n_i}}{|\vec r - \vec r'|} d\vec r' + \frac{\partial}{\partial n_i} \left ( \frac{\delta E_{xc}}{\delta\rho(\vec r)} \right )\\
\end{split}
\label{apx:eq-derivada_Janak-02}
\end{equation}

But, how to calculate $ \frac{\partial}{\partial n_i} \left ( \frac{\delta E_{xc}[\rho]}{\delta \rho(\vec r)} \right ) $? Let $\rho = \sum_i n_i \rho_i $ and $\rho_i = |\phi_i|^2$, we reach

\begin{equation}
\begin{split}
\frac{\partial}{\partial n_i} \left ( \frac{\delta E_{xc}[\rho]}{\delta \rho(\vec r)} \right ) &= \int \frac{\partial}{\partial \rho(\vec r)}\left ( \frac{\delta E_{xc}[\rho]}{\delta \rho(\vec r)} \right ) \frac{\partial \rho}{\partial n_i} d\vec r  \\
                                                                                               &= \int \frac{\delta^2 E_{xc}}{\delta\rho(\vec r) \underbrace{\delta\rho(\vec r')}_{\text{$ = \rho_i + \sum_j n_j \frac{\partial \rho_j}{\partial n_i} $}}}  d\vec r'
\end{split}
\label{apx:eq-derivada_Janak-03}
\end{equation}

Therefore,

\begin{equation}
\begin{split}
\frac{\partial (V_H + v_{xc})}{\partial n_\alpha} &= \int \frac{\rho_i(\vec r)}{|\vec r - \vec r'|} d\vec r' + \int \frac{\sum_j n_j\frac{\partial \rho_j(\vec r)}{\partial n_i}}{|\vec r - \vec r'|} d\vec r' +   \int \frac{\delta^2 E_{xc}}{\delta\rho(\vec r)} \left [ \rho_i(\vec r) + \sum_j n_j \frac{\partial \rho_j}{\partial n_i} \right ]  d\vec r'      \\
\end{split}
\label{apx:eq-derivada_Janak-04}
\end{equation}

But, from Hellmann-Feynman theorem, $ \frac{\partial F}{\partial \lambda} = \int \psi^*_{\lambda} \frac{\partial H_\lambda}{\partial\lambda}\psi_\lambda d\vec r $, and then

\begin{equation}
\begin{split}
 \frac{\partial (V_H + v_{xc})}{\partial n_\alpha} &= \int\int \frac{\rho_i(\vec r)\rho_i(\vec r')}{|\vec r - \vec r'|} d\vec r d\vec r' + \int\int \frac{\sum_j n_j\frac{\partial \rho_j(\vec r)}{\partial n_i}}{|\vec r - \vec r'|} \rho_i(\vec r') d\vec r d\vec r' \\
                                             &+   \int\int \frac{\delta^2 E_{xc}}{\delta\rho(\vec r)} \rho_i(\vec r') \left [ \rho_i(\vec r) + \sum_j n_j \frac{\partial \rho_j}{\partial n_i} \right ]  d\vec r d\vec r'  .    \\
\end{split}
\label{apx:eq-derivada_Janak-05}
\end{equation}

\end{document}